\documentclass[11pt]{article}
\usepackage{moriond,epsfig}

\def\be{\begin{equation}}
\def\ee{\end{equation}}
\def\bea{\begin{eqnarray}}
\def\eea{\end{eqnarray}}

\begin{document}
\vspace*{4cm}
\title{Photon Structure and Heavy Flavour Production in 
\boldmath $\gamma\gamma$ Collisions at LEP}

\author{Armin B\"ohrer}

\address{FB Physik, Siegen University, Emmy-Noether-Campus,\\
57068~Siegen, Germany}
\maketitle\abstracts{
New results reported at the Moriond-QCD 2003 conference 
from the experiments ALEPH, DELPHI, L3, and OPAL on the structure
of the photon as well as heavy quark production in $\gamma\gamma$
collisions are presented. The hadronic structure of the photon is 
studied in single-tagged events; the interaction of two virtual 
photons is investigated in double-tagged events. Di-jet events
allow, besides constraining the photon structure, investigations on
the structure of jets, measurement of the di-jet as well as inclusive and
differential cross sections. Inclusive
charm and bottom production are investigated at LEP~2 energies.
The total and differential cross sections for charm quarks are now
measured by all four LEP collaborations, the total bottom by two.
Charmonia are detected inclusively via the muonic decay of the J$/\psi$
and separated for the resolved and diffractive processes. Updated results
are available for exclusive production of the $\eta_{\mathrm c}$ meson.
The search for exclusive $\eta_{\mathrm b}$ production is presented.}

\section{Introduction}
\begin{picture}(10,0)(0,0)
\put(405.,520.){SI-2003-2}
\put(405.,505.){May 2003}
\end{picture}
The hadronic photon structure function 
is measured in single-tagged events. The structure function separates into 
two parts, the pointlike part, which is calculated in perturbative QCD and 
the hadronlike part, which is not accessible in perturbative QCD at low 
$x$. Here, when the photon fluctuate into a state similar to a light 
vector meson, the gluon content of the photon is important. 
Double-tagged events test the interaction of two virtual photons and 
probe next-to-leading QCD predictions. These topologies might restrict BFKL 
calculations, where the DGLAP evolution is not applicable. The study of 
di-jet events gives insight into the structure of the photon as well. The 
jets may carry a certain fraction $x$ of the momentum of the 
incoming photon. In the direct process, when the two photons produce a 
quark anti-quark pair, both jets carry large $x$. For single and double 
resoved processes, when one or both photon appear resolved, the fractions 
may be smaller. Di-jet events offer even a richer spectrum of measurements 
and tests, such as differential cross section, jet shapes etc. 

Inclusive heavy flavour production in two-photon collisions is dominated 
by two processes, the direct and single-resolved process. It therefore 
reveals the structure of the photon and is sensitive to its gluon content. 
At LEP~2 energies direct and the single-resolved processes contribute in 
equal shares to heavy flavour final states. The charm cross section is 
about two orders of magnitude larger than the bottom production due to 
the smaller quark mass and higher electric charge. The large quark mass allows 
the production of heavy flavour to be calculated in perturbative QCD, 
where the resolved part also depends on the assumed gluon density of the 
photon.

The exclusive charmonium production has a diffractive contribution at 
low $p_{\mathrm T}^2$ of the vector meson (pomeron exchange) and a resolved 
contribution at high $p_{\mathrm T}^2$ (gluon exchange). 
The resolved production 
of J$/\psi$, when calculated in the nonrelativistic QCD, predicts that 
the colour-octet contribution dominates. Exclusive charmonia and 
bottomonia production at LEP~2 provides a precise tool to test QCD 
at low energies. Their two-photon widths and masses are constraint by 
approaches used in lattice QCD, nonrelativistic QCD and potential models.

In this article we summarize the progress made with respect to last 
year on the photon structure and in heavy flavour production in two-photon 
collisions at LEP: improvements, updates and new results are reviewed. 
The inclusive hadron and pair production are discussed in a separate 
talk~\cite{lin} at this conference. 
For a prior status of two-photon physics, we recommend Ref.~\cite{ascona}; 
for results mentioned in this report but not cited also refer to 
Ref.~\cite{ascona} and references therein. A short and recent 
more general overview on two-photon physics and the interaction 
of the photon can be found in Ref.~\cite{sumascona}.

\section{Photon Structure}

\subsection{Hadronic Photon Structure Function}
Two new results are contributed to this conference, one final 
publication~\cite{f2gammaA} (ALEPH) and a preliminary 
result~\cite{f2gammaD} (DELPHI). 
The photon with the larger momentum transfer $Q^2$, which is radiated by 
the beam-electron with the larger scattering angle probes the target photon's 
structure. The latter is radiated as a real photon such that the scattered 
beam-electron escapes undetected down the beam-pipe. From the measurement 
of the invariant mass $W_{\gamma\gamma}$ of the hadronic system produced 
by the two colliding photons $x = Q^2 / (Q^2+W_{\gamma\gamma}^2)$ can 
be calculated: Then $x$ is 
the momentum fraction of the parton in the real photon, which is 
probed. (See Ref.~\cite{ascona,nisius} for a general overview on photon 
structure functions.) 
New results are consistent with previous measurements at similar $Q^2$ 
as shown in Figure~\ref{figf2gamma}. The data disfavour structure 
function with large gluon 
content, which predict a steep rise versus low $x$. 

It has been noticed that with the improved understanding and larger 
statistics due to the LEP measurements a value for the strong coupling 
constant is extracted, competitive and compatible with other 
measurements~\cite{alphas}.

\subsection{Double-Tagged Events}
The interaction of two virtual photons is invesigated using 
double-tagged events. In the new contribution~\cite{doubleA} from the 
ALEPH collaboration, which is being published, the measurements are 
compared to Monte-Carlo predictions, a next-to-leading order 
(NLO) calculations and to the BFKL approach. The data 
are consistent with previous measurements. Monte-Carlo models describe the 
shape of various distributions of the data quite well. The NLO calculation 
predicts a cross section lower than is measured, while the shape follows 
the data in most distributions. Little room is left for further 
contributions, especially at high $Y$, with $Y=\ln(s_{\mathrm{ee}}/s_0) 
\approx W_{\gamma\gamma}^2/\sqrt{Q_1^2 Q_2^2}$ (Where this approximation 
requires that the invariant mass of the hadronic system 
$W_{\gamma\gamma}$ is larger that the individual momentum transfers 
$Q_i$, i.e., $W_{\gamma\gamma}^2 \gg Q_i^2$.): 
LO-BFKL is therefore ruled out. NLO-BFKL is in reasonable 
agreement with the measurements.

\subsection{Di-Jet Production}
The study of di-jet production by OPAL in collisions of two quasi-real 
photons~\cite{dijetO} is a rich laboratory both for the investigation 
of the structure of the photon, jet properties and physics associated 
with jet production in general. Only a small part can be indicated here. 
From the energies and momenta of the hadrons and jets, the data sample 
can be separated into direct, single- and double-resolved processes 
calculating $x_{\gamma}$, like $x$ before, an estimate of the 
momentum fraction of the parton in the real photon. 
The inclusive di-jet cross section as function of the mean transverse 
jet energy extracted is nicely described by 
NLO, when PYTHIA or HERWIG is used for hadronization correction. 
Deviations are only seen for the double-resolved events. This might be 
a hint for multiple interactions. Jet-profiles are reproduced by Monte-Carlo 
models. The angle between the di-jet axis and the incoming partons 
shows evidence for the quark and gluon exchange contributions. The first 
measurement of the differential cross section of function of $x_{\gamma}$ 
is shown. Again disagreement is obserbved for double-resolved 
events: the difference of data and NLO prediction is just as large as 
the contribution predicted by the model for multiple interaction of the 
partons in the photon.

\section{Heavy Flavour Production}

\subsection{Inclusive D$^{*\pm}$ Production}
All four LEP experiments measure now the inclusive charm 
production using most of their LEP~2 statistics (corresponding to an 
integrated luminosity of $\approx 700\,{\mathrm{pb^{-1}}}$) at 
energies around $\sqrt{s} \approx 200\,{\mathrm{GeV}}$ 
with fully reconstructed D$^{*\pm}$ mesons in no-tag events; a new publication 
from ALEPH 
is contributed to Moriond~\cite{DstarA} (Also added in the total cross 
section diagramm is the update of L3~\cite{chabotL} using leptons as 
charm tag.), see Figure~\ref{figcc}.

The experiments provide differential 
distributions in pseudorapidity. The distributions are found to be 
flat in this variable, what is in agreement with the expectation 
for NLO calculations. The distribution in transverse 
momentum to the beam axis as predicted in NLO calculations in the 
massive approach agrees with the data. (More recent 
calculations~\cite{spies} using massive or massless charm quarks 
also describe the data well.)

Direct and single-resolved contribution can be separated using the fact that
in the resolved one the remnant jet carries away a part of the invariant
mass available in the $\gamma\gamma$ collision. 
The relative contribution fitted with the data of the ALEPH 
experiment are found in agreement 
with the NLO prediction and with previous measurements. 

\subsection{Total Charm Cross Section}
\begin{figure}
\begin{minipage}{.45\textwidth}
~ \\[7mm]
\epsfxsize=1.05\textwidth
\epsfysize=1.05\textwidth
\begin{picture}(10,0)(0,0)
\put(212.,12.){$x$}
\put(0.,192.){$F_{2}^{\gamma}/\alpha$}
\end{picture}
\epsfbox{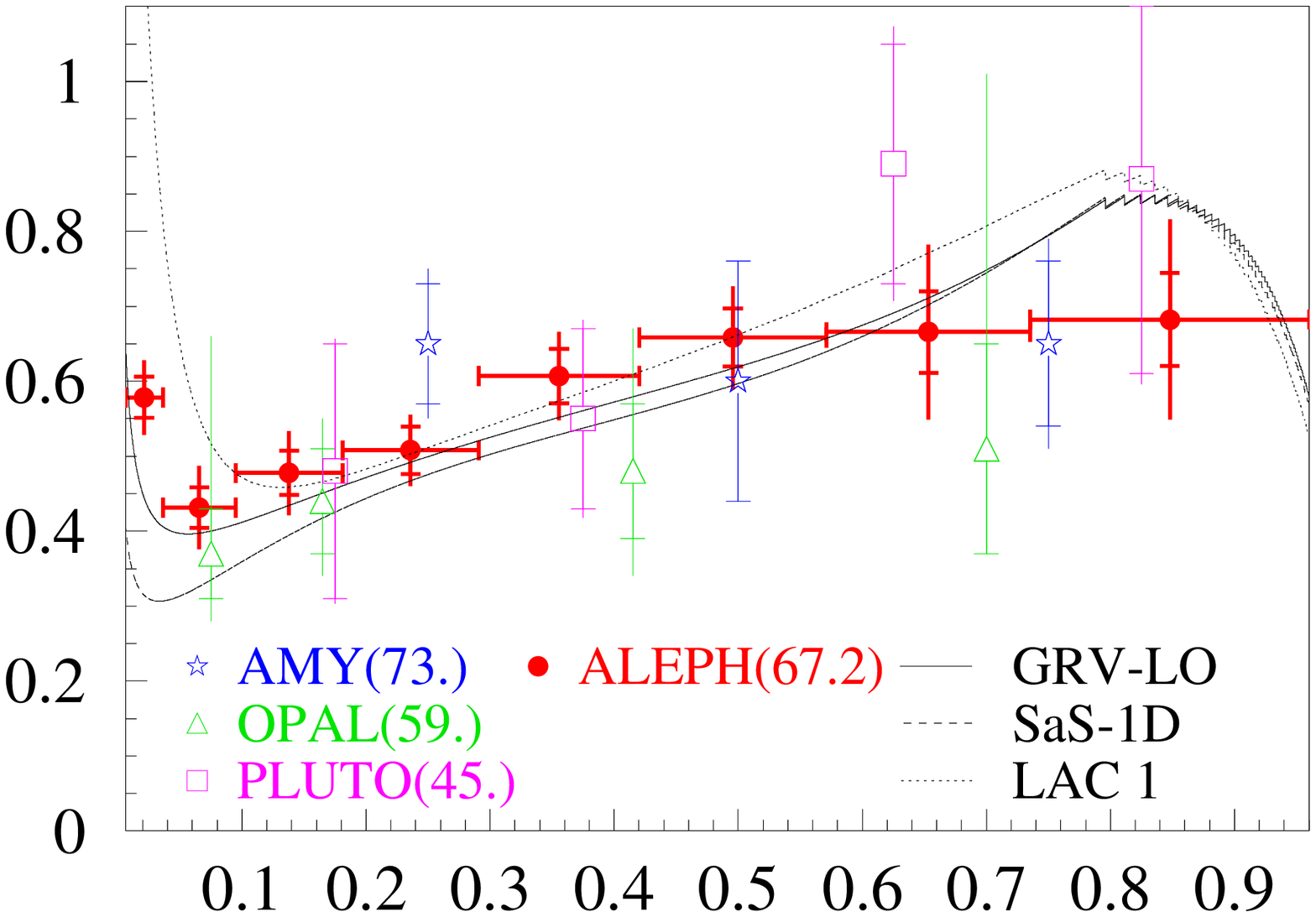}
\end{minipage}
\begin{minipage}{.08\textwidth}
~
\end{minipage}
\begin{minipage}{.45\textwidth}
\epsfxsize=1.00\textwidth
\epsfysize=1.00\textwidth
\epsfbox{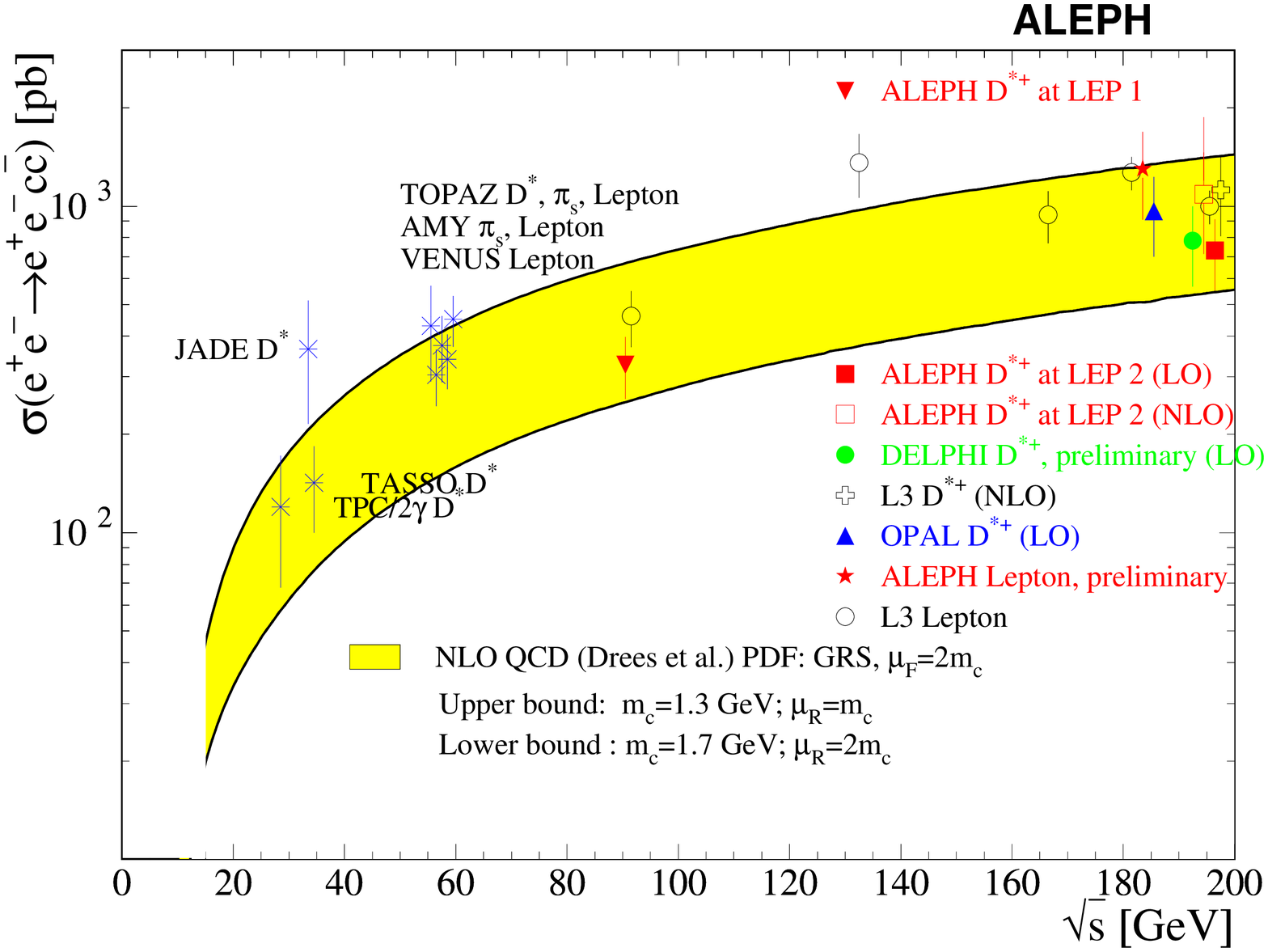}
\end{minipage}

\begin{minipage}{.45\textwidth}
\caption{Hadronic structure function for $Q^2$ around $16\,$GeV$^2$. 
\label{figf2gamma}}
\end{minipage}
\begin{minipage}{.08\textwidth}
~
\end{minipage}
\begin{minipage}{.45\textwidth}
\caption{Inclusive charm cross section at various e$^+$e$-$ colliders.
\label{figcc}}
\end{minipage}

\end{figure}

Extrapolated to the full phase space, the measurements using the D$^{*\pm}$ as 
the charm tag, can be compared to NLO QCD calculations and other 
measurements, see Figure~\ref{figcc}. All data are consistent. 
If only the direct contribution is considered the 
prediction at LEP~2 would be lower by a factor two. It should be 
noted that for the measurements with leptonic final state a light charm 
quark mass is slightly preferred.

\subsection{Charm Structure Function $F_{{\mathrm c},2}^{\gamma}$}
When one of the scattered beam particles is detected, the event can be used
to determine the charm structure function $F^2_{\gamma,{\mathrm {c}}}$. With
$55.3 \pm 11.0$ such single-tagged events with a ${\mathrm{D}}^{*\pm}$ meson 
from the full LEP~2 statistics, the OPAL
collaboration published a measurement in two bins of $x$ with $\langle 
Q^2 \rangle \approx 20\,{\mathrm{GeV}}$~\cite{chaF2gO}. 
The comparison with the calculations
shows that a point-like contribution is not sufficient to describe the data.
A hadron-like part is needed. The data even exceed the models, though the
measurement errors are still too large to be conclusive.

\subsection{Inclusive Bottom Production}
Open bottom production is measured by the L3 and the OPAL
col\-la\-bo\-ra\-tions at LEP~2 energies. For this conference 
L3 updates their result to the full 
integrated luminosity at LEP~2~\cite{chabotL}. 
Their analysis procedures exploit the
fact, that the momentum as well as the transverse momentum of leptons
with respect to the closest jet is higher for muons and electrons from bottom
than from background, which is mainly charm. Therefore, leptons with momenta
of more than $2\,{\mathrm{GeV}}$ are selected and their momentum distribution
with respect to the closest jet is investigated.

The total cross section measurements for open bottom production
are compared to NLO calculations, showing that the calculations 
underestimate the data by a factor three corresponding to three to four 
standard deviations.

\subsection{Inclusive J$/\psi$}
DELPHI investigates the inclusive J$/\psi$ production in 
$\gamma\gamma$ collisions at LEP~2 energies~\cite{JpsiD} with a 
clean signal of $36\pm 7$ muonic J$/\psi$ decays. 
From a fit of the diffractive 
and resolved contribution, as taken from the PYTHIA simulation, a value 
of $74 \pm 22\%$ is extracted to originate from the resolved process: a
clear indication for the gluon in the photon. In a recent paper the octet 
production of J$/\psi$ in association with jets has been 
discussed~\cite{JpsiKla} and found to be needed 
for an agreement with the DELPHI measurements. 

\subsection{Exclusive $\eta_{\mathrm c}$}
The formation of the $\eta_{\mathrm c}$ in exclusive production in 
two-photon production is a good test of QCD (See Ref.~\cite{braccini} for 
a short summary of the present status of studies of exclusive particle 
production in two-photon events). A preliminary study of 
$\eta_{\mathrm c}$ production at LEP~2 is contributed by 
the DELPHI collaboration to this 
conference~\cite{etacD}. Nice signals are found in the decay modes 
of the $\eta_{\mathrm c}$ to $\pi^+\pi^-$K$^+$K$^-$, K$^+$K$^-$K$^+$K$^-$, 
and K$_{\mathrm {S}}$K$^+\pi^-$. No signal, however, is seen in the 
$\pi^+\pi^-\pi^+\pi^-$ decay mode, though it is expected when 
the branching fraction of Ref.~\cite{RPP} are implied. An upper limit on 
the two-photon width for this channel is given, while for the other three 
a value somewhat higher than the world average~\cite{RPP} and recent 
measurements~\cite{braccini} is extracted.

\subsection{Exclusive $\eta_{\mathrm b}$}
As reported at last year's Moriond conference, 
the ALEPH experiment published a search for the still undiscovered
$\eta_{\mathrm b}$ pseudoscalar meson in $\gamma\gamma$ collisions via 
exclusive production, i.e., all decay products are observed in the 
detector. A preliminary search from L3~\cite{etabL} is presented 
here. 

The mass of the $\eta_{\mathrm b}$ can be extracted, e.g., 
from potential models, pQCD,
NRQCD, and lattice calculations. While the production can reliably be
estimated the branching ratios of the meson are guessed from an 
MLLA combined with LPHD and using isospin invariance~\cite{etabascona}. 
Including detector efficiency and acceptance about one to two signal 
events are expected; background events about one. 

The meson has been search for in four different decay channels by L3. 
Three candidates are found (while ALEPH had found one 
candidate, but with a better mass resolution), which are compatible 
with the background expectation. Limits are given for the product 
of the particle width times branching ratio of a few hundred eV. 

\section{Summary}
In this article we summarize the ten contributions on the structure 
of the photon and 
of heavy flavour production in two-photon collisions submitted to 
the Moriond-QCD 2003 conference. The space being limited, we recommend 
the reader to consult the papers in the bibliography and references therein 
for the very details. 


\end{document}